\documentclass[12pt]{iopart}
\usepackage{iopams}

\usepackage{graphicx}   
\usepackage{verbatim}   
\usepackage{color}      
\usepackage{hyperref}   

\newcommand{\cP}{\mathcal{P}}
\newcommand{\cT}{\mathcal{T}}
\newcommand{\cK}{\mathcal{K}}

\begin{document}

\title[${\cal PT}$ Electronics]{${\cal PT}-$symmetric electronics}

\author{J. Schindler, Z. Lin, J. M. Lee, H. Ramezani, F. M. Ellis, T. Kottos}

\address{Department of Physics, Wesleyan University, Middletown, CT 06459, USA}
\ead{zlin@wesleyan.edu}

\begin{abstract}
We show both theoretically and experimentally that a pair of inductively coupled active LRC circuits (dimer), one with amplification 
and another with an equivalent amount of attenuation, display all the features which characterize a wide class of non-Hermitian systems 
which commute with the joint parity-time ${\cal PT}$ operator: typical normal modes, temporal evolution, and scattering processes.
Utilizing a Liouvilian formulation, we can define an underlying ${\cal PT}$-symmetric Hamiltonian, which provides important insight
for understanding the behavior of the system. When the ${\cal PT}$-dimer is coupled to transmission lines, the resulting scattering 
signal reveals novel features which reflect the ${\cal PT}$-symmetry of the scattering target. Specifically we show that the device 
can show two different behaviors simultaneously, an amplifier or an absorber, depending on the direction and phase relation of the 
interrogating waves. Having an exact theory, and due to its relative experimental simplicity, ${\cal PT}$-symmetric electronics offers 
new insights into the properties of ${\cal PT}$-symmetric systems which are at the forefront of the research in mathematical 
physics and related fields.
\end{abstract}


\section{Introduction}

Among the many recent developments in ${\cal PT}$ systems, the application of pseudo-Hermitian ideas into the realm of 
electronic circuitry not only promises a new generation of electronic structures and devices, but also provides a platform 
for detailed scrutiny of many new concepts within a framework of easily accessible experimental configurations. A first 
example of this was the demonstration in Ref.~\cite{SLZEK11} that a pair of coupled $LRC$ circuits, one with 
amplification and the other with equivalent amount of attenuation, provided the simplest experimental realization of a 
${\cal PT}$ symmetric system. With a normal mode structure where all dynamical variables are easily measured in the time 
domain, extensions of the circuit approach will provide a valuable testing ground for further developments into more 
sophisticated structures. Moreover, the ${\cal PT}$-circuitry approach suggested also opens new avenues for innovative 
electronics architectures for signal manipulation from integrated circuits to antenna arrays, and allows for direct contact 
with cutting edge technological problems appearing in (nano)-antenna theory, split-ring resonator arrays, and meta-materials. 

Examples of ${\cal PT}$-symmetric systems range from quantum field theories and mathematical physics \cite{BB98,BBM99,BBJ02,BBJM07} 
to atomic \cite{HKO06,GKN08}, solid state \cite{BFKS09,BFKS09b,JSBS10} and classical optics \cite{MGCM08,L09,GSDMVASC09,RMGCSK10,RKGC10,ZCFK10,
L10b,LREKCC11,SXK10}. A ${\cal PT}$-symmetric system can be described by a phenomenological "Hamiltonian" ${\cal H}$ which may 
have a real energy spectrum, although in general ${\cal H}$ is non-Hermitian. Furthermore, as some parameter $\gamma$ that controls the 
degree of non-Hermiticity of ${\cal H}$ changes, a spontaneous ${\cal PT}$ symmetry breaking occurs. At this point, $\gamma=
\gamma_{\cal PT}$, the eigenfunctions of ${\cal H}$ cease to be eigenfunctions of the ${\cal PT}$-operator, despite the fact 
that ${\cal H}$ and the $\mathcal{PT}$-operator commute \cite{BB98}. This happens because the ${\cal PT}$-operator is anti-linear, 
and thus the eigenstates of ${\cal H}$ may or may not be eigenstates of ${\cal PT}$. As a consequence, in the {\it broken} 
$\cal{PT}$-symmetric phase the spectrum becomes partially or completely complex. The other limiting case where both ${\cal H}$ 
and ${\cal PT}$ share the same set of eigenvectors corresponds to the so-called {\it exact} $\mathcal{PT}$-symmetric phase in 
which the spectrum is real. This result led Bender and colleagues to propose an extension of quantum mechanics based on non-
Hermitian but ${\cal PT}$-symmetric operators \cite{BB98,BBM99}. The class of non-Hermitian systems with real spectrum has 
been extended by other researchers in order to include Hamiltonians with generalized ${\cal PT}$ (antilinear) symmetries \cite{M03}.

While the applicability of these ideas in the quantum framework is still being debated, optical systems provide a particularly 
fertile ground where ${\cal PT}$-related concepts can be realized \cite{MGCM08} and experimentally investigated \cite{GSDMVASC09,
RMGCSK10}. In this framework, ${\cal PT}$ symmetry demands that the complex refractive index obeys the condition $n({\vec r})=n^*(-{\vec r})$. 
${\cal PT}$-synthetic materials can exhibit several intriguing features. These include among others, power oscillations and 
non-reciprocity of light propagation \cite{MGCM08,RMGCSK10,ZCFK10}, absorption enhanced transmission \cite{GSDMVASC09}, and 
unidirectional invisibility \cite{LREKCC11}. Despite these efforts and the consequent wealth of theoretical results associated 
with ${\cal PT}$-structures, until very recently only one experimental realization of a system with balanced gain and loss has been reported 
\cite{RMGCSK10}. These authors studied the light propagation in two coupled ${\cal PT}$ symmetric waveguides where the spontaneous 
${\cal PT}$-symmetry breaking ``phase transition'' \cite{note1} was indirectly confirmed. The analysis relied on the paraxial 
approximation which under appropriate conditions maps the scalar wave equation to the Schr\"odinger equation, with the axial 
wavevector playing the role of energy and with a fictitious time related to the propagation distance along the waveguide axis. 

This observation led us recently to propose a new set-up based on active LRC circuits where the novel features of ${\cal PT}$-
symmetric structures can reveal themselves and can be studied both theoretically and experimentally in great detail. The system 
consists of a 
pair of coupled electronic oscillators, one with gain and the other with loss. This "active" dimer, is implemented with simple
electronics, and allow not only for a direct observation of a spontaneous ${\cal PT}$-symmetric ``phase transition'' from a 
real to a complex eigenfrequency spectrum but also for its consequences in the spatio-temporal domain. At the same time the 
equivalent scattering system, where a localized ${\cal PT}$ symmetric structure is connected to one or two transmission line 
(TL) leads allow us to access the validity of recent theoretical predictions \cite{L10b,L10d,CGS11,LRKCC11,S10,CDV07,M09,CGCS10}. 

This paper presents our recent results pertaining to the ${\cal PT}$ electronics. We begin with a general discussion of electronics 
in the context of ${\cal PT}$ symmetric systems in section~\ref{sec:elec}. Then in section~\ref{sec:dim_modes} we examine the normal 
mode structure of the simplest such circuit, the ${\cal PT}$ dimer. We experimentally demonstrate how it displays all the novel phenomena 
encountered in systems with generalized ${\cal PT}$-symmetries. Section~\ref{sec:dim_dyn} discusses the unique aspects of ${\cal PT}$ 
dynamics exhibited by the dimer, particularly upon passage from the exact to broken phase. In section~\ref{sec:scat} we investigate 
the simplest possible scattering situation where the dimer is coupled to a single TL, and derive a non-unimodular conservation relation 
connecting the left and right reflectances. A direct consequence of this relation is the existence of specific frequencies for which 
the system behaves either as a perfect absorber or as an amplifier, depending on the side (gain or loss) to which the TL is coupled. 
In section~\ref{sec:scat2} we demonstrate theoretically and experimentally that a two-port ${\cal PT}$-symmetric electronic cavity 
can act as a simultaneous coherent perfect absorber (CPA)-amplifier. Our circuit is the electronic equivalent of a CPA-Laser 
device which was recently proposed in the optics framework, and constitutes the first experimental realization of such devices. 
Finally, section~\ref{sec:expcon} presents several issues involving practical implementation of ${\cal PT}$ circuits, along with 
some related experimental details.  Our conclusions are given in section~\ref{sec:concl}.


\section{${\cal PT}$ electronics}\label{sec:elec}

One of the most convenient advantages of an electronic approach is that, at least in the low frequency domain, where the wavelength is significantly greater than the dimensions of the circuit, all spatial symmetry considerations can be reduced to a matter of network topology defined through the application of Kirchoff's laws. Physical symmetry is irrelevant as long as the network has the desired node topology and the connecting elements are appropriately valued. Analogous to the familiar case of a ${\cal PT}$-symmetric potential, the parity operation is equivalent to the interchange of labels corresponding to pairs of associated circuit components. 

\begin{figure}
\center
\includegraphics[scale=0.5]{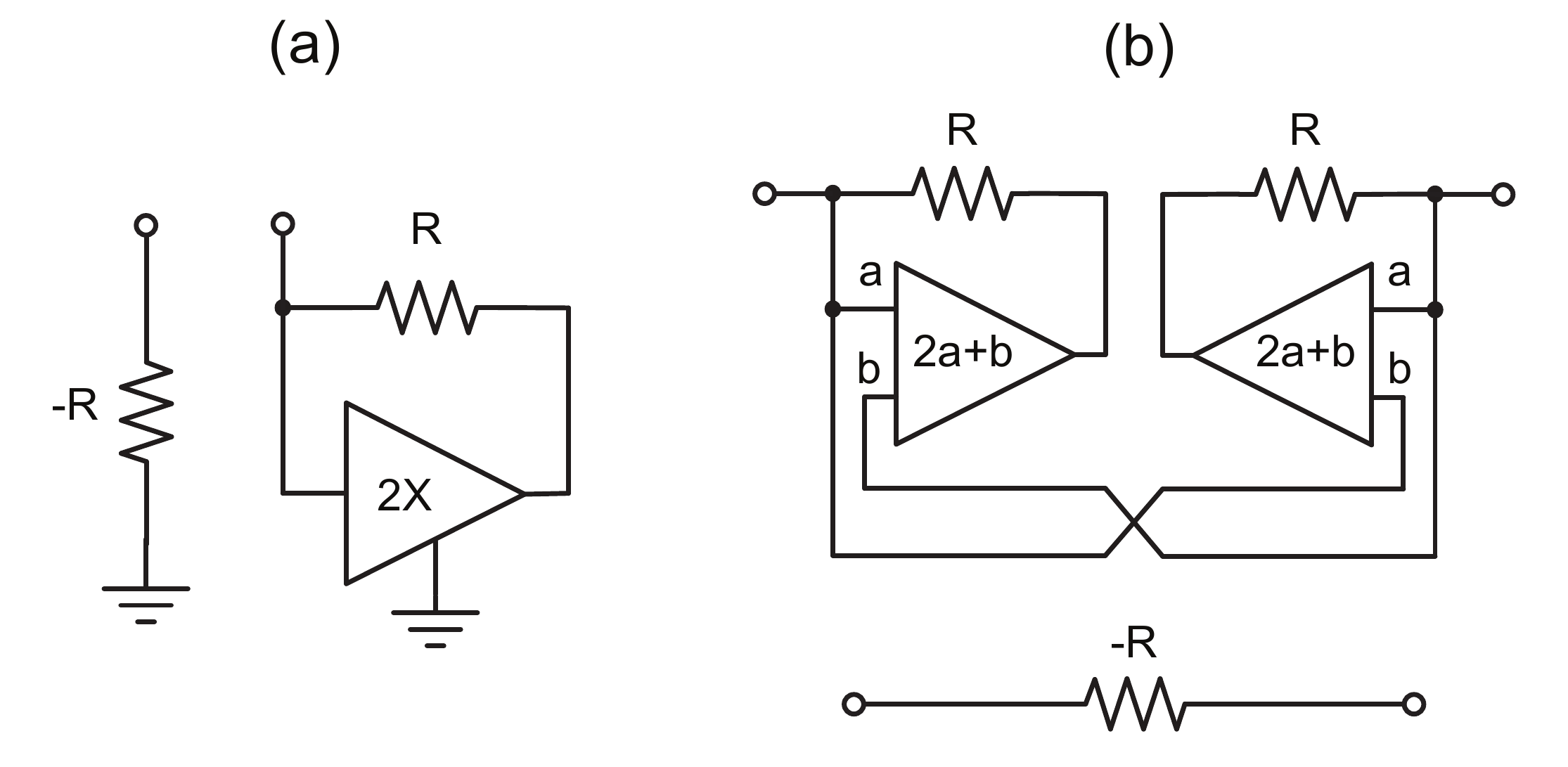}
  \caption{Two negative impedance converters with their equivalents. (a) A ground referenced negative resistance node. (b) A floating, two-terminal negative resistance.}
\label{fig:nics}
\end{figure}

For simplicity, we will restrict our discussion to the usual fundamental physical devices: resistors, capacitors and inductors. Only the resistor, due to it's dissipative nature, requires modification upon time-reversal where we include generic Ohmic elements with either positive or negative resistance. Negative resistance represents the simplest conceptual inclusion of amplification into electronics since Kirchoff's laws can be used without modification. Fig.~\ref{fig:nics} illustrates how simple linear amplifiers can be configured to achieve negative resistance. The schematic implementation in (a) results in a single, ground-referenced node, while that in (b) shows a true two-terminal configuration. The former is of greatest utility due to it's simplicity and the pervasiveness of ground nodes (defining a common zero potential) in typical circuits. Section~\ref{sec:expcon} discusses further details of the experimental negative resistance converters.

\begin{figure}
\centering
\includegraphics[width=3.5in]{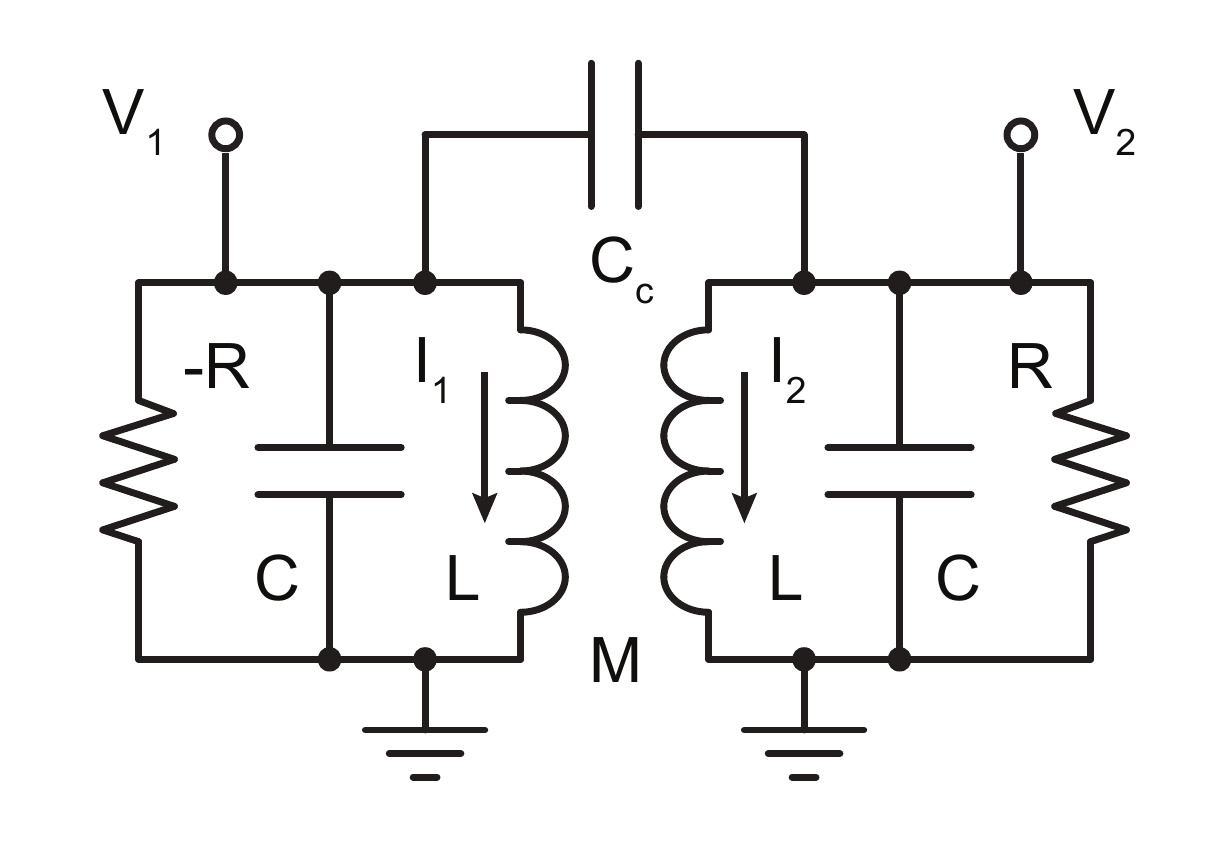}
  \caption{Schematic of the ${\cal PT}$-symmetric electronic dimer. Both mutual inductance coupling and capacitive coupling are included for generality.}
\label{fig:basic_dimer}
\end{figure}

Theoretical analysis of circuits including negative resistance elements, however, requires respecting a subtle condition: any two terminal circuit structure reducing to a pure negative resistance will be undefined unless the structure is placed in parallel with a capacitance. This conclusion results from the divergence of the pole associated with parallel $RC$ combinations (with negative $R$) in the limit of $C \rightarrow 0$. This pole only arises if parallel capacitance is initially considered, so is hidden, and often overlooked in the consideration of negative resistance circuits. It is inconsequential with normal, positive resistance, where it's sign corresponds to exponential decay. For example, the standard series $LRC$ circuit, though it appears to have a mathematically well behaved solution for negative $R$, is invalid in that realm due to the hidden pole. Our choice of the parallel $LRC$ configurations was dictated by this consideration.

Thus, for a ${\cal PT}$-symmetric circuit incorporating these basic elements, it is necessary that (1) all reactive elements either have representation in parity-associated network pairs, or connect parity inverted network nodes, (2) all Ohmic elements are paired with opposite sign, and (3) each negative Ohmic element has an associated parallel capacitance, or AC equivalent, as part of the circuit. Valid ${\cal PT}$-circuits of arbitrary complexity can be built up using these simple rules, though their stability needs to be independently determined. In principle, the long-wave approximation could be relaxed with an appropriate inclusion of waveguide connections, however, this would return geometry into the mix of ${\cal PT}$ considerations.


\section{${\cal PT}$ dimer modes}\label{sec:dim_modes}

Figure~\ref{fig:basic_dimer} shows the ${\cal PT}$-symmetric dimer, the simplest configuration with a non trivial (more 
than one mode) pseudo-Hermitian spectrum. Both capacitive and mutual inductive coupling are included for generality, although 
the experimental results presented throughout this work are exclusively one or the other. The gain side on the left of 
Fig.~\ref{fig:basic_dimer} is indicated by $-R$ and was implemented using the configuration of Fig.~\ref{fig:nics}(a). 
The loss on the right is achieved with a conventional resistance of the same value, resulting in the gain/loss parameter 
$\gamma = R^{-1} \sqrt{L/C}$ for this system. Further details of the experimental circuit are given in section.~\ref{sec:expcon}.

Kirchoff's laws for the dimer with both mutual inductance coupling and capacitive coupling between the oscillators are given 
for the gain side (Eq.~(\ref{kirch1})) and loss side (Eq.~(\ref{kirch2})).
\begin{eqnarray}
V_1=i\omega(LI_1+MI_2) \quad\quad I_1-\frac{V_1}{R}+i\omega CV_1+i\omega C_c(V_1-V_2)=0  \label{kirch1}\\
V_2=i\omega(LI_2+MI_1) \quad\quad I_2+\frac{V_2}{R}+i\omega CV_2+i\omega C_c(V_2-V_1)=0 \label{kirch2}
\end{eqnarray}
Eliminating the currents from the relations, scaling frequency and time by $\omega_0=\sqrt{L/C}$, and taking $\mu = M/L$ 
and $c = C_c / C$ gives the matrix equation:
\begin{equation}
\left(
\begin{array}{cc}
\frac{1}{\omega(1-\mu^2)}-\omega(1+c)-i\gamma & \omega c-\frac{\mu}{\omega(1-\mu^2)} \\
\omega c-\frac{\mu}{\omega(1-\mu^2)} & \frac{1}{\omega(1-\mu^2)}-\omega(1+c)+i\gamma
\end{array}
\right)
\left(
\begin{array}{c}
V_1 \\
V_2
\end{array}
\right)=0.
\end{equation}
At this point, it is obvious that the system is ${\cal PT}$ symmetric: swapping the indices and changing the sign of $i$ 
leaves the equations unchanged. This linear, homogeneous system has four normal mode frequencies, as required to fulfill 
any arbitrary initial condition for voltage and current, given by
\begin{equation}
\hspace{-2cm}\omega_{1,2}=\pm\frac{\sqrt{\gamma_{c}^2-\gamma^2}+\sqrt{\gamma_{PT}^2-\gamma^2}}{2\sqrt{1+2c}};\quad 
\omega_{3,4}=\pm\frac{\sqrt{\gamma_{c}^2-\gamma^2}-\sqrt{\gamma_{PT}^2-\gamma^2}}{2\sqrt{1+2c}};
\label{equ:omegk}
\end{equation}
with the ${\cal PT}$ symmetry breaking point identified as
\begin{equation}
\gamma_{PT}=|\frac{1}{\sqrt{1-\mu}}-\sqrt{\frac{1+2c}{1+\mu}}|
\end{equation}
and the upper critical point by
\begin{equation}
\gamma_{c}=\frac{1}{\sqrt{1-\mu}}+\sqrt{\frac{1+2c}{1+\mu}}.
\end{equation}
Note that the given forms explicitly show all of the relationships among the critical points and the real and imaginary 
parts of the frequencies. The exact phase, $0 < \gamma < \gamma_{PT}$, is characterized by four purely real eigenfrequencies 
coming in two pairs of positive ($\omega_1,\omega_3>0$) and negative ($\omega_2,\omega_4<0$) values, while in the 
broken phase below the upper critical point, $\gamma_{PT} < \gamma < \gamma_{c}$ the eigenfrequencies are coming in complex
conjugate pairs with non-vanishing real parts, and above $\gamma_{c}$, as two purely imaginary complex conjugate pairs. 
The broken phase of the ${\cal PT}$ dimer is unstable, in that it is ultimately dominated by an exponentially growing mode. 

The normal modes in the exact phase are characterized by equal magnitudes for the voltage oscillations in the gain and 
loss sides, which in the $+\omega$, real part convention allowed by the real eigenfrequencies, are given by
\begin{equation}
\left(
\begin{array}{c}
V_1 \\
V_2
\end{array}
\right)_\pm=\frac{1}{\sqrt{2}}
\left(
\begin{array}{c}
1 \\
-\exp(i\phi_\pm)
\end{array}
\right)
\end{equation}
with a phase $\phi_\pm$ of the loss side
\begin{equation}
\phi_\pm=\pi/2-\tan^{-1}\left[\frac{1}{\gamma}\left(\frac{1}{(1-\mu^2)\omega_\pm}-(1+c)\omega_\pm\right)\right].
\label{equ:phipm}
\end{equation}
As the gain/loss parameter traverses the exact region, $0\leq\gamma\leq\gamma_{PT}$, the phase progresses from the in- 
and out-of-phase configuration of a Hamiltonian coupled oscillator, to a mode coalescence at $\gamma_{PT}$ with $\phi_\pm 
\sim \pi/2$ with the real frequency
\begin{equation}
\omega_+=\omega_-=\left[(1-\mu^2)(1+c)\right]^{-1/4}
\end{equation}

Examination of the inductor currents,
\begin{equation}
\label{equ:curr}
\left(
\begin{array}{c}
I_1 \\
I_2
\end{array}
\right)_{\pm}
=
\left(
\begin{array}{cc}
\frac{1}{1-\mu^2} & -\frac{\mu}{1-\mu^2} \\
-\frac{\mu}{1-\mu^2} & \frac{1}{1-\mu^2}
\end{array}
\right)
\left(
\begin{array}{c}
V_1 \\
V_2
\end{array}
\right)_{\pm}
\end{equation}
reveal phase shifts, relative to the corresponding voltages, that advance on the gain side and retard on the loss side 
within either mode. This is as required for the net transfer of electrical energy from the gain side to the loss side 
as the gain/loss parameter increases. This evolutionary behavior is helpful in understanding the spectral and dynamical 
behavior of the dimer.

An alternate analysis of the dimer is also accomplished by recasting Kirchoff's laws, Eqs.~(\ref{kirch1}) and Eq. (\ref{kirch2}) 
into a ``rate equation" form by making use of a Liouvillian formalism
\begin{equation}
\label{liuvilian1}
{d{\bf \Psi}\over d\tau} = {\cal L}  {\bf \Psi};\quad 
{\cal L}= \left(
\begin{array}{cccc}
 0 & 0 & 1 & 0 \\
 0 & 0 & 0 & 1 \\
 -\alpha  \beta  & \alpha  \zeta  & (1+c) \gamma  & c \gamma  \\
 \alpha  \zeta  & -\alpha  \beta  & -c \gamma  & -(1+c) \gamma
\end{array}
\right)
\end{equation}
where $\alpha=1/(1-\mu^2)$, $\beta=1+c+c\mu$, $\zeta=c+\mu+c\mu$ and ${\bf \Psi}\equiv (Q_1, Q_2, {\dot Q_1}, {\dot Q_2})^T$ 
with $Q_n=CV_n$. This formulation opens new exciting directions for applications \cite{rsegk12} 
of generalized ${\cal PT}$-mechanics \cite{M03} as it can be interpreted as a Schr\"odinger equation with non-Hermitian 
effective Hamiltonian $H_{\rm eff}=i{\cal L}$. This Hamiltonian is symmetric with respect to generalized ${\cal P}_0{\cal 
T}_0$ transformations \cite{rsegk12}, i.e. $[{\cal P}_0{\cal T}_0,H_{\rm eff}]=0$, where
\begin{equation}
\label{generaltransf}
{\cal P}_0=\left ( \begin{array}{cc}
\sigma_x & 0\\
0        &\sigma_x
 \end{array} \right);\quad
{\cal T}_0=\left ( \begin{array}{cc}
{\bf 1} & 0\\
0        &-{\bf 1}
 \end{array} \right) {\cal K}
\end{equation}
and $\sigma_x$ is the Pauli matrix, ${\bf 1}$ is the $2\times 2$ identity matrix, and ${\cal K}$ denotes the operation
of complex conjugation. By a similarity transformation ${\cal R}$ \cite{rsegk12}, 
\begin{equation}
\hspace{-2cm}{\cal R}=\left(
\begin{array}{cccc}
 \frac{2 i (b+d)}{1+\sqrt{1+2 c}} & -\frac{2 i (b+d)}{1+\sqrt{1+2 c}} & \frac{-1-2 c+\sqrt{1+2 c}}{c} & \frac{-1-2 c+\sqrt{1+2 c}}{c} \\
 \frac{i \left(-1-2 c+\sqrt{1+2 c}\right) (b-d)}{c} & \frac{i \left(-1-2 c+\sqrt{1+2 c}\right) (b-d)}{c} & \frac{2}{1+\sqrt{1+2 c}} & -\frac{2}{1+\sqrt{1+2 c}} \\
 \frac{i \left(1+2 c-\sqrt{1+2 c}\right) (b-d)}{c} & \frac{i \left(1+2 c-\sqrt{1+2 c}\right) (b-d)}{c} & \frac{2}{1+\sqrt{1+2 c}} & -\frac{2}{1+\sqrt{1+2 c}} \\
 \frac{2 i (b+d)}{1+\sqrt{1+2 c}} & -\frac{2 i (b+d)}{1+\sqrt{1+2 c}} & \frac{1+2 c-\sqrt{1+2 c}}{c} & \frac{1+2 c-\sqrt{1+2 c}}{c}
\end{array}
\right)
\end{equation}
$H_{\rm eff}={\cal R}^{-1}H {\cal R}$ can be related to a transposition symmetric, $\cP\cT-$symmetric Hamiltonian $H=H^T=
\cP H^\dagger \cP$, $\cT=\cK$ where $\cP={\cal R}P_0{\cal R}^{-1}$. The matrix $H$ is then 
\begin{equation}
\label{liuvilian2}
 H=-\left(
\begin{array}{cccc}
 0 & b+i r & d+i r & 0 \\
 b+i r & 0 & 0 & d-i r \\
 d+i r & 0 & 0 & d-i r \\
 0 & d-i r & d-i r & 0
\end{array}
\right)
\end{equation}
where $b=\sqrt{\alpha  \left(\beta +\sqrt{\beta ^2-\zeta ^2}\right)/{2}}$, $d=\sqrt{\alpha  \left(\beta -\sqrt{\beta ^2-
\zeta ^2}\right)/{2}}$ and $r=\frac{1}{2}\sqrt{1+2 c} \gamma$. The frequencies and normal modes within this framework are 
identical to Eqs.~(\ref{equ:omegk},\ref{equ:curr}).

These normal mode properties can be measured in our electronic dimer by simultaneous observation of the node voltages $V_1$ and $V_2$ of Fig.~\ref{fig:basic_dimer}. Our set-up allows detailed analysis for gain/loss parameters $\gamma$ on either side of the ${\cal PT}$-phase transition point. In the exact phase, time series samples are captured with the dimer slightly unbalanced to marginally oscillate the mode of interest. Beyond the critical point, a transient sample is obtained dominated by the exponentially growing mode. Details are given in section~\ref{sec:expcon}. 

\begin{figure}
\centering
\includegraphics[scale=0.35]{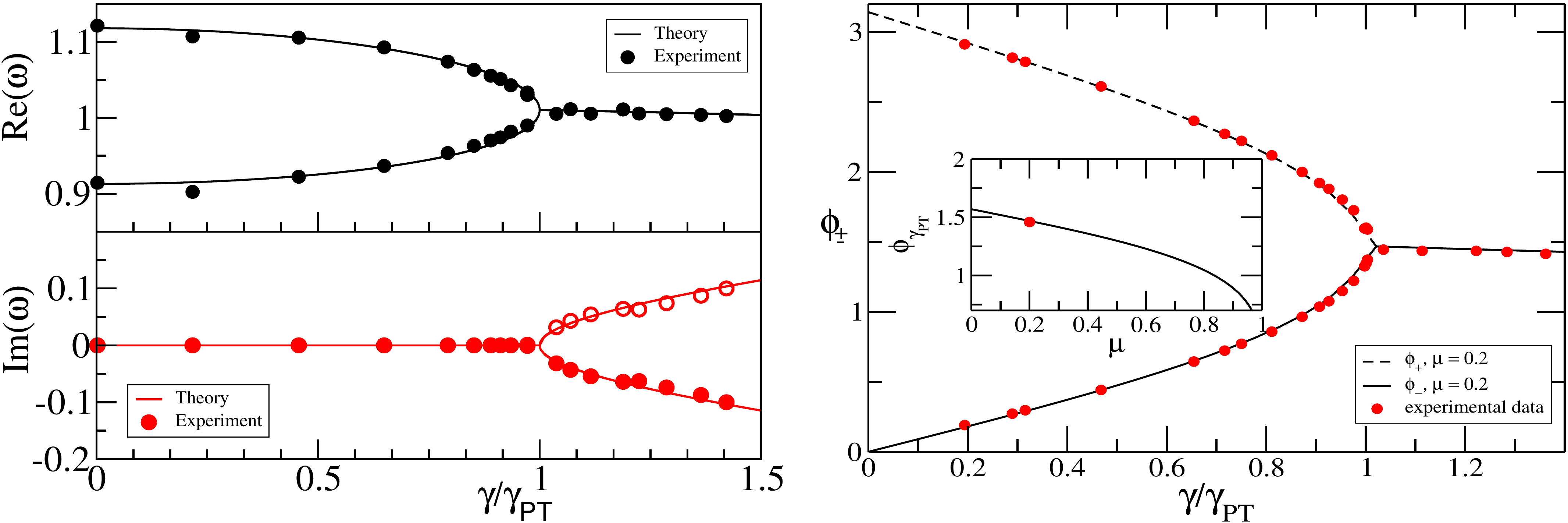}
\caption{(Left) Parametric evolution of the experimentally measured eigenfrequencies, vs. the normalized gain/loss parameter 
$\gamma/\gamma_{\cal PT}$. A comparison with the theoretical results of Eq.~(\ref{equ:omegk}), indicates an excellent agreement. 
In all cases, we show only the ${\cal R}e (\omega_{l})>0$ eigenfrequencies. The open circles in the lower panel are reflections 
of the experimental data (lower curve) with respect to the ${\cal I}m(\omega)=0$ axis. (Right) Parametric evolution of the 
phase difference $\phi_{\pm}$. Symbols correspond to experimental data while the lines indicate the theoretical results of 
from Eq.~(\ref{equ:phipm}). The theoretical $\phi_{\cal PT}(\mu)$ is shown in the inset.}
\label{fig:exp_modes}
\end{figure}

In Fig.~\ref{fig:exp_modes} we report measurements for the dimer frequencies (left) and inter-component phases (right) 
compared with the theoretical expressions, Eq.~(\ref{equ:omegk}) and Eq.~(\ref{equ:phipm}) respectively, for the values 
$\mu=0.2$ and $c=0$. The ${\cal PT}$ symmetry imposes the condition that the magnitude of the two voltage components are 
equal to one-another in the exact phase. This property is also experimentally observed. For $\gamma=0$, the phases 
corresponding to the symmetric and antisymmetric combination are $\phi_{-}=0$ and $\phi_{+}=\pi$, respectively. When 
$\gamma$ is subsequently increased and the system is below the ${\cal PT}$ threshold, the eigenstates are not orthogonal 
and their phases can be anywhere (depending on $\gamma/\gamma_{\cal PT}$) in the interval $[0,\pi]$.

The value of phase difference at the spontaneous ${\cal PT}$-symmetric breaking point $\gamma=\gamma_{\cal PT}$ can be calculated analytically and it is given by the expression:
\begin{equation}
\label{phicr}
\phi_{\cal PT}(\mu)=
\arccos\left(\frac{\sqrt{1-\sqrt{1-\mu^2}}}{\sqrt{1+\sqrt{1-\mu^2}}}\right)
\end{equation}
We note that in the limit of $\mu\rightarrow 0$ we get $\phi_{\cal PT}=\pi/2$, corresponding to a "circular" polarization of the eigenmode. 
The opposite limit of $\mu\rightarrow 1$ results to $\phi_{\cal PT}=0$ corresponding to "linear" polarization.


\section{${\cal PT}$ dimer dynamics}\label{sec:dim_dyn}

The signatures of ${\cal PT}$-symmetry and the transition from the exact phase to the
broken phase are similarly reflected in the temporal behavior of our system. Eq.~(\ref{liuvilian1}) can be solved either analytically 
or via direct numerical integration in order to obtain the temporal behavior of the capacitor charge $Q_n(\tau)$ and the displacement 
current $I_n(\tau)$ in each of the two circuits of the ${\cal PT}$-symmetric dimer. As an example of the dimer state evolution, we 
consider an initial displacement current in one of the circuits with all other dynamical variables zero. 

\begin{figure}
\includegraphics[width=\hsize,keepaspectratio]{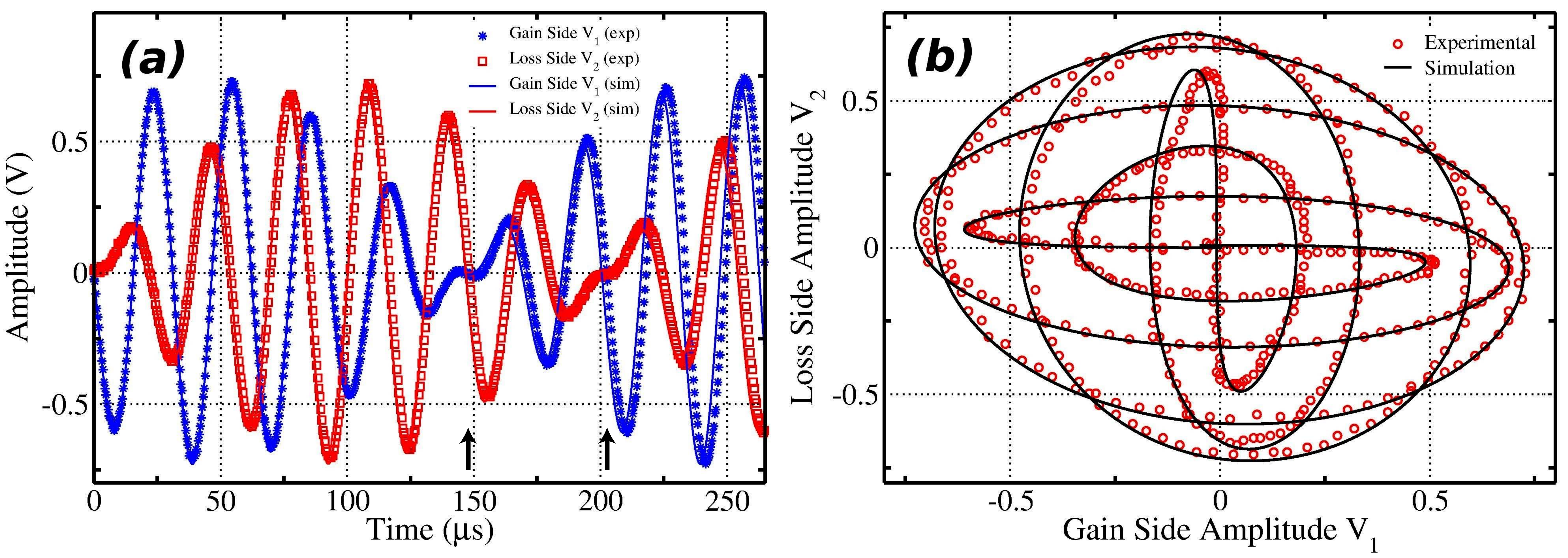}
\caption{
(a) Gain and loss side voltages vs. time compared to the simulation. (b) Gain vs. loss side Lissajous figure for one 
beat period. At $t=0$ an initial current was imposed in the gain side inductor with all other dynamical variables zero. Note that the end of the beat (indicated by the arrow near $200 \mu s$) is preceded by a similar point where both voltages pass through zero (indicated by the arrow near $150 \mu s$) with $V_2$ decreasing, and $V_1$ stationary. This corresponds to the complementary initial condition starting from the loss 
side, and illustrates an asymmetric time between the beat nodal points of oscillatory activity in the two oscillators of 
the dimer.
}

\label{fig:beats}
\end{figure}

In Fig.~\ref{fig:beats} we present some typical measurements for the temporal behavior of circuit voltages along with the 
corresponding numerical result. We consider a dimer configuration with $\mu=0.2$ and $c=0$ (i.e. inductive coupling only). 
In the left panel of Fig.~\ref{fig:beats}a, we show $V_1(\tau)$ and $V_2(\tau)$ for an initial condition having $I_1(0) =1.2 mA$ 
with all other dynamical variables zero. The right panel shows the same data as a Lissajiou plot, with the initial condition 
trajectory leaving the origin with $V_1$ decreasing, and $V_2$ stationary. Agreement between the experiment (circles) and the simulations (lines) is observed, illustrating that, in spite of the presence of dissipative elements and non-orthogonal states, the beat superposition associated with real frequencies occurs. There is, however, a subtle distinction: Since energy is not conserved, the beat is asymmetric between the gain side and the loss side nodal times, with oscillatory activity spending more time between gain side nodal points as energy grows to a significantly larger size before decaying and growing between the loss side nodal points. However, unlike traditional coupled-oscillator beats, instead of ``slashing" between both sides during the course of the beats, a growth and decay energy dance occurs with both sides more or less equally represented except in the vicinity of the nodal points. This behavior is a direct result of the non-orthogonal phase relationships that become more pronounced as $\gamma \rightarrow \gamma_{PT}$. A Hamiltonian dimer would exhibit a perfect half-beat offset between the left and right voltage beat envelopes.

\begin{figure}
\centering
\includegraphics[width=4in,keepaspectratio,clip]{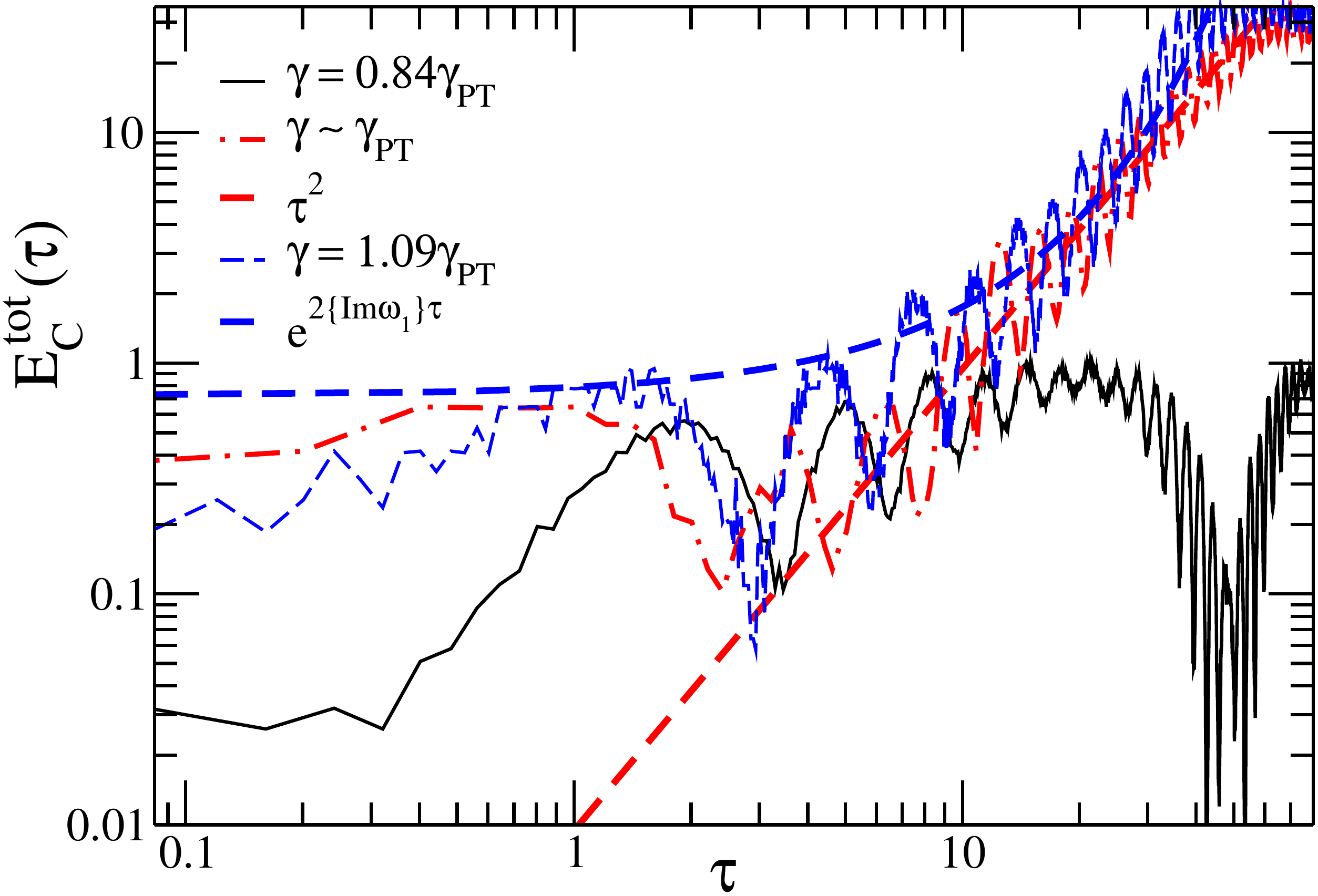}
\caption{Experimentally measured temporal dynamics of the capacitance energy $E_C^{\rm tot}(\tau)$ of the total system for various $\gamma$-values. As $\gamma\rightarrow \gamma_{\cal PT}$ the $\tau^2$ behavior signaling the spontaneous ${\cal PT}$-symmetry breaking is observed.}
\label{fig:temp_dyn}
\end{figure}

We have also traced these energy dance features by studying the time-dependence of the total capacitance energy:
\begin{equation}
\label{Ectot}
E_C^{\rm tot}(\tau)= {Q_1^2(\tau)+Q_2^2(\tau) \over 2C}.
\end{equation}
With the initial condition used in the experiment, we expect power oscillations which are due to the unfolding of the non-orthogonal eigenmodes \cite{BB98,MGCM08,ZCFK10,RMGCSK10}. This universal feature is evident in the temporal behavior of $E_C^{\rm tot}(\tau)$ as can be seen in Fig. \ref{fig:temp_dyn}. On the other hand, for $\gamma> \gamma_{\cal PT}$ the dynamics is unstable and $E_C^{\rm tot}(\tau)$ grows exponentially with a rate given by the maximum imaginary eigenvalue $\max\{{\cal I}m(\omega_{l})\}$ (see Fig. \ref{fig:temp_dyn}).

The most interesting behavior appears at the spontaneous ${\cal PT}$-symmetry breaking point $\gamma=\gamma_{\cal PT}$. At this point the matrix ${\cal L}$ has a defective eigenvalue. In this case, the evolution $U=\exp({\cal L}\tau)$ can be calculated from the Jordan decomposition of ${\cal L}$ as ${\cal J}=S{\cal L}S^{-1}$. Having in mind the form of the exponential of a Jordan matrix, it follows immediately that linear growing terms appear in the evolution of the charge vector $\left(Q_1(\tau), Q_2(\tau)\right)^T$ \cite{H10}. This results in a quadratic increase of the capacitance energy i.e. $E_C^{\rm tot}(\tau)\sim \tau^2$. Although all systems typically becomes very sensitive to parameters near a critical point, we are able to control the circuit elements sufficiently well to observe the approach to the predicted $\tau^2$ behavior of the energy. This time range is limited by the dynamic range of our circuit linearity, as discussed in section~\ref{sec:expcon}


\section{The Janus faces of ${\cal PT}$-symmetric scattering}\label{sec:scat}

\begin{figure}
\includegraphics[scale=0.5]{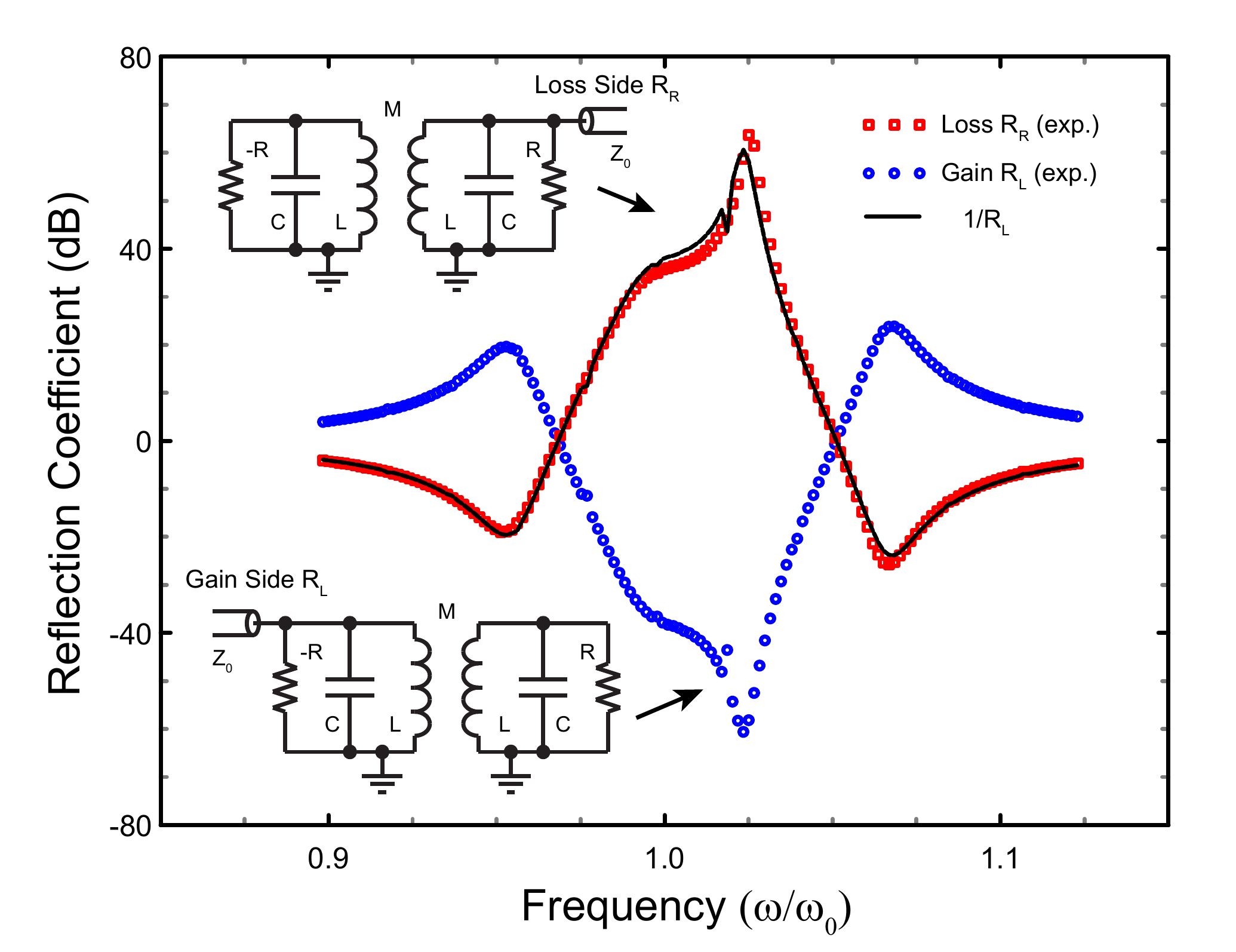}
\centering
\caption{(Color) 
Two experimental configurations associated with a simple ${\cal PT}$-symmetric dimer. In the lower and upper
circuits, we couple a transmission line to the gain and loss sides, respectively. Preliminary experimental measurements for
the corresponding reflection coefficients are shown (loss-side red, gain-side blue) along with the solid line corresponding to
$R_{\rm L}^{-1}$ illustrate the reciprocal nature $R_{\rm L} R _{\rm R}=1$ (see text) of the ${\cal PT}$-scattering. Here $\mu=0.2$,
$\gamma=0.164$ and $Z_0=15.5\sqrt{L/C}$.
}
\label{fig:reflect1}
\end{figure}

We report our initial scattering studies with the following two reciprocal geometries: In the first case, a transmission line (TL) is 
attached to the left (amplified) circuit of the dimer load while in the second case, the TL is connected to the right (lossy) circuit 
of the load (see lower and upper insets of Fig. \ref{fig:reflect1} respectively). Experimentally, the equivalent of a TL with 
characteristic impedance $Z_0$ could be attached to either side of the dimer at the $RLC$ circuit voltage node in the form of a resistance 
$R_0=Z_0$ in series with a variable frequency voltage source. The right and left traveling wave components associated with the TL would 
be deduced from the complex voltages on both sides of $R_0$. With $V_{LC}$ the voltage on the $LC$ circuit, and $V_0$ the voltage on the 
synthesizer side of the coupling resistor $R_0$, the right (incoming) wave has a voltage amplitude $V_L^{+}=V_0/2$ and the left (reflected) 
wave has a voltage amplitude $V_L^{-}=V_{LC}-V_0/2$. The voltage source defines the phase of the incoming wave.

At any point along a TL, the current and voltage determine the amplitudes of the right and left traveling wave components \cite{LSEK12}. The 
forward $V_{L,R}^{+}$ and backward $V_{L,R}^{-}$ wave amplitudes, and $V_{1,2}$ and $I_{1,2}$ the voltage and current at the left or right 
TL-dimer contacts satisfy the continuity relation
\begin{equation}
\label{scatstates}
V_{1,2}=V_{L,R}^{+}+V_{L,R}^{-};\quad I_{1,2}=\left[V_{L,R}^{+}-V_{L,R}^{-}\right]/Z_0 
\end{equation}
which connect the wave components to the currents and voltages at the TL-dimer contact points. Note that with this convention, a positive 
lead current flows into the left circuit, but out of the right circuit, and that the reflection amplitudes for left or right incident 
waves are defined as $r_L\equiv {V_L^{-}/V_L^{+}}$ and $r_R\equiv {V_R^{+}/V_R^{-}}$ respectively.

Application of Kirchoff's laws at the TL leads allow us to find the corresponding wave amplitudes and reflection. For this analysis, we assume the $e^{-i\omega t}$ wave convention. For example, the 
case of the left-attached lead in the lower inset of Fig. \ref{fig:reflect1} gives
\begin{eqnarray}
\label{Kirchoff}
\eta (V_L^{+}-V_L^{-})= I_1^{M} - \gamma V_1 - i\omega V_1 \quad \quad  \nonumber \\
V_1 = -i\omega \left[ I_1^M + \mu I_2^M\right]; \quad V_2 = -i\omega \left[ I_2^M + \mu I_1^M\right] \\
0 = I_2^{M} + \gamma V_2 - i\omega V_2 \quad \quad \nonumber
\end{eqnarray}
where $\eta=\sqrt{L/C}/Z_0$ is the dimensionless TL impedance, and $I_{1,2}^M$ are the current amplitudes in the left or right inductors. 
These are equivalent to the simple dimer form Eq.~(\ref{kirch1}) and Eq.~(\ref{kirch2}) with the addition of the contact current and the 
opposite sign convention for $i$ more appropriate for the traveling wave analysis. Similar equations apply for the right-attached case 
shown in the upper inset of Fig. \ref{fig:reflect1}. We are interested in the behavior of the reflectance $R_{L/R}\equiv |r_{L/R}|^2$, 
as the gain/loss parameter $\gamma$, and the frequency $\omega$ changes. 

For ${\cal PT}$-symmetric structures, the corresponding scattering signals satisfy {\it generalized unitarity relations} which reveal 
the symmetries of the scattering target. Specifically, in the single-port set up this information is encoded solely in the reflection. 
To unveil it, we observe that the lower set-up of Fig. \ref{fig:reflect1} is the ${\cal PT}$-symmetric replica of the upper one. Assuming 
therefore that a potential wave at the left lead (lower inset) has the form $V_L(x)=\exp(ikx)+r_L\exp(-ikx)$ (we assume $V_L^+=1$ and 
$V_L^-=r_L$ in Eq. (\ref{scatstates})), we conclude that the form of the wave at the right lead associated with the upper circuit of 
Fig. \ref{fig:reflect1} is $V_R(x)=\exp(-ikx)+r_r\exp(ikx)= V_L^*(-x)$. Direct comparison leads to the relation
\begin{equation}
\label{r_Lr_R}
r_L\cdot r_R^* = 1 \rightarrow R_L= 1/R_R \quad {\rm and} \quad \phi_L=\phi_R
\end{equation}
where $\phi_{L/R}$ are the left/right reflection phases. Note that Eq. (\ref{r_Lr_R}) differs from the more familiar conservation 
relation $R=1$, which applies to unitary scattering processes as a result of flux conservation. In the latter case left and right 
reflectances are equal. Instead in the ${\cal PT}$-symmetric case we have in general that $R_L\neq R_R$ \cite{A01}. 

For the specific case of the ${\cal PT}$-symmetric dimer, we can further analytically calculate the exact expression for the reflection 
coefficients. From Eqs. (\ref{Kirchoff}) we have
\begin{eqnarray}
\label{r_R}
r_L(\omega)=-f(-\eta, -\gamma)/f(\eta, -\gamma) \quad \quad \nonumber \\
r_R(\omega)=-f(-\eta, \gamma)/f(\eta, \gamma) \quad \quad \quad \\
f=1-\left[2-\gamma m(\gamma+\eta)\right]\omega^2+m\omega^4-i\eta \omega (1-m\omega^2) \nonumber \\
{\rm with} \quad m=1/\sqrt{1-\mu^2} \quad \quad \quad \nonumber
\end{eqnarray}
In the limiting cases of $\omega\rightarrow 0,\infty$ the reflection amplitude becomes $r_R\rightarrow \mp 1$ and thus unitarity is restored.

In the main panel of Fig. \ref{fig:reflect1} we plot the reflection coefficients of  Eq. (\ref{r_R}) for the two scattering configurations 
shown in the sub-panels. The measured reflectances $R_L$, and $R_R$ satisfy the generalized conservation relation $R_L\cdot R_R=1$ as expected
from Eq. (\ref{r_Lr_R}). The slight deviation from reciprocity in the vicinity of large reflectances can be attributed 
to nonlinear effects. 

A peculiarity of our results is the appearance of a singularity frequency point $\omega_J(\mu,\gamma)$ for which $R_R\rightarrow \infty$, while 
a reciprocal point for which $R_L=0$ is also evident. The corresponding $(\omega_{\rm J};\gamma_{\infty,0})$ are found from Eq. (\ref{r_R}) to 
be 
\begin{equation}
\label{LA}
\gamma_{\infty,0}={1\over 2} \left( \sqrt{\eta^2+{4\mu^2\over (1-\mu^2)}}\mp\eta\right);\, \omega_{\rm J}={1\over \sqrt{1-\mu^2}} 
\end{equation}
Therefore, our experiment demonstrates that a ${\cal PT}$-symmetric load is a simple electronic Janus device that for the same values of 
the parameters $\omega_J,\mu,\gamma$ acts as a perfect signal absorber as well as a signal amplifier, depending on the side (gain or loss) that
the TL is coupled to the dimer. 

For the more general case of a two-port ${\cal PT}$ scattering, it was shown theoretically in \cite{CGS11} and later on confirmed experimentally in 
\cite{LSEK12} that the following conservation relation holds: 
\begin{equation}
\label{TR_scat}
\sqrt{R_LR_R}=|T-1|
\end{equation}
Equation (\ref{r_Lr_R}) is a special case of Eq. (\ref{TR_scat}) once we realize that in the single port case the transmittance $T=0$.

\section{Two-port coherent perfect absorber-amplifier}\label{sec:scat2}

Recent theoretical studies in the optics framework \cite{L10b} have suggested that a two-port ${\cal PT}$-symmetric cavity 
can act as a simultaneous coherent perfect absorber (CPA)-laser. In this section we provide the first experimental 
realization of this proposal using  a two-port configuration of our ${\cal PT}$-symmetric electronic dimer, and demonstrate it's 
action as simultaneous CPA-amplifier. We consider the capacitively coupled case, with $c=C_c/C$ as previously 
defined, to demonstrate the independence of the generic behavior from the coupling mechanism. Following steps similar to the single-port case, 
Kirchoff's laws lead to the following set of equations:
\begin{eqnarray}
\label{bigscaledeqL}
\hspace{-1cm}\eta(V_{L}^{+} - V_{L}^{-}) + (V_{L}^{+} + V_{L}^{-}) \left[ i\omega (1+c) + \frac{1}{i\omega} + \gamma \right] - i\omega c(V_{R}^{+} + V_{R}^{-}) =0 \\
\hspace{-1cm}-\eta(V_{R}^{+} - V_{R}^{-}) + (V_{R}^{+} + V_{R}^{-}) \left[ i\omega (1+c) + \frac{1}{i\omega} - \gamma \right] - i\omega c(V_{L}^{+} + V_{L}^{-}) =0  
\end{eqnarray} 

The above equations can be written in a more elegant form by making use of the transfer matrix formulation:
\begin{eqnarray}
\left(
\begin{array}{c}
V_{R}^{+} \\
V_{R}^{-}
\end{array}
\right) 
&= \mathcal{M}
\left(
\begin{array}{c}
V_{L}^{+} \\
V_{L}^{-}
\end{array} 
\right);\quad
\mathcal{M} &= \frac{1}{2 \omega c \eta }
\left(
\begin{array}{cc}
A + iB & iC \\
-iD & A - iB
\end{array}
\right)  
\label{mmatrixgeneral} 
\end{eqnarray} 
where the transfer matrix elements $\mathcal{M}$ are 
$A = 2\eta \Omega $, 
$B = \Omega^{2} - \eta^2 - \omega^2 c^2 + \gamma^2$,
$C = (\gamma - \eta)^2 + \Omega^2 - \omega^2 c^2$, 
and $D = (\gamma + \eta)^2 + \Omega^2 - \omega^2 c^2$,
with $\Omega = \omega (1+c) -1/\omega$. 
One can further express the spectral transmission and reflection coefficients for left (L) and right (R) 
incidence in terms of the transfer matrix elements as \cite{CDV07,M09}
\begin{equation}
\label{trans_refl}
t_L=t_R\equiv t = {1\over {\mathcal M}_{22}},\quad r_L=-{{\mathcal M}_{21}\over {\mathcal M}_{22}},\quad r_R={{\mathcal M}_{12}\over {\mathcal M}_{22}}
\end{equation}
where we have used the identity that $\det(\mathcal{M}) = 1$. 

\begin{figure}
\centering
\includegraphics[scale=0.4]{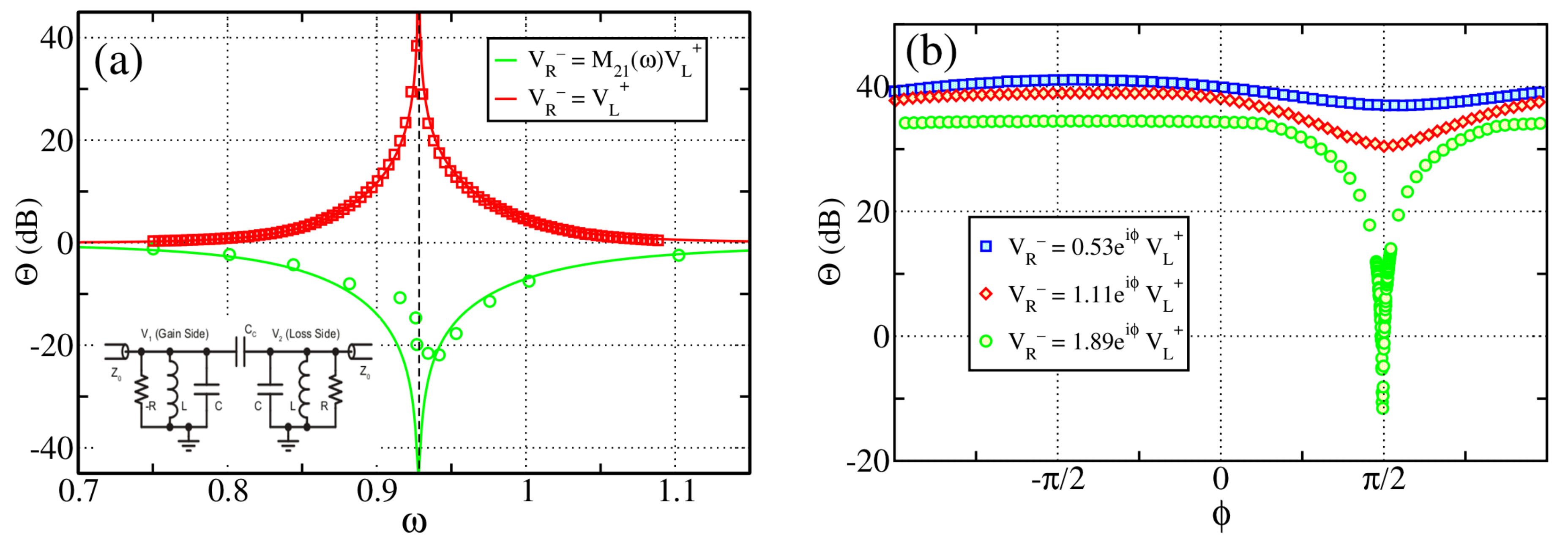}
  \caption{(Color) (a) The overall output coefficient $\Theta(\omega)$ around the Janus amplification/attenuation frequency $\omega_{J}$ (vertical dashed line)
for a ${\cal PT}$-symmetric electronic circuit coupled to two ports.
The parameters used in this simulation are $\eta=0.110$, $\gamma=0.186$ and $c=0.161$. The red curve corresponds to the two port coherent
input excitation with $V_{R}^{-} = \mathcal{M}_{21}(\omega)V_{L}^{+}$; the blue curve correspond to a two-port input signal with 
$V_{R}^{-} = V_{L}^{+}$. In the former case the system acts as an perfect attenuator while in the latter as an amplifier. The dots are experimental values. (b) Plots of experimental $\Theta(\omega_J)$ as the loss side input excitation phase is changed, for several excitation amplitudes. Note the extremely sharp dependence at the Janus condition.}
\label{fig:lasabs}
\end{figure}

An alternative formulation of the transport problem utilizes the so-called scattering matrix $\mathcal{S}$ which connects incoming 
to outgoing waves and its elements can be written in terms of the transmission/reflection coefficients. Specifically 
\begin{eqnarray}
\left(
\begin{array}{c}
V_{R}^{+} \\
V_{L}^{-}
\end{array} 
\right) 
&= \mathcal{S}
\left(
\begin{array}{c}
V_{L}^{+} \\
V_{R}^{-}
\end{array} 
\right);\quad\quad
\mathcal{S} = \frac{1}{\mathcal{M}_{22}}
\left(
\begin{array}{cc}
1 & \mathcal{M}_{12} \\
-\mathcal{M}_{21} & 1
\end{array}
\right)
\label{smatrixgeneral}
\end{eqnarray} 
Using the scattering matrix language one can derive conditions in order our ${\cal PT}$-symmetric structure to act either as
an amplifier or as a perfect absorber. For a laser oscillator without an injected signal, the boundary conditions $V_{L}^{+} = 
V_{R}^{-}=0$ apply, which imply from Eq. (\ref{smatrixgeneral}) the condition $\mathcal{M}_{22}(\omega) = 0$ \cite{M09}. In 
contrast, for a perfect absorber the boundary conditions, $V_{L}^{-} = V_{R}^{+}=0$, corresponding to zero reflected waves, 
hold. From Eq.~(\ref{smatrixgeneral}) this implies $\mathcal{M}_{11}(\omega) = 0$, while the amplitudes of the incident waves 
must satisfy the 
condition $V_{R}^{-} = \mathcal{M}_{21} (\omega)V_{L}^{+}$. In general, the condition for an amplifier/laser system, is not
satisfied simultaneously with the condition for a perfect absorber. However for any ${\cal PT}$-symmetric structure, one can show
from Eq.~(\ref{mmatrixgeneral}) that the matrix elements of ${\mathcal M}$ satisfy the relation $\mathcal{M}_{22}(\omega) = 
\mathcal{M}^{*}_{11}(\omega^{*})$ \cite{L10b}. As a result, a real $\omega = \omega_{J}$ exists, that satisfies the amplifier/laser 
condition simultaneously with the absorber condition ($\mathcal{M}_{22} (\omega_{J}) = \mathcal{M}_{11}(\omega_{J}) = 0$). 
Hence the two-port ${\cal PT}$-symmetric dimer can behave simultaneously as a perfect absorber and as an amplifier. This property 
can be explored using an overall output coefficient $\Theta$ defined as \cite{L10b}
\begin{equation}
\Theta = \frac{|V_{R}^{+}|^{2} + |V_{L}^{-}|^{2}}{|V_{L}^{+}|^{2} + |V_{R}^{-}|^{2}} \label{thetadef}
\end{equation}
Note that in the case of a single-port scattering set-up discussed earlier in this section, the $\Theta$-function collapses to
the left/right reflectances. We can further simplify the above expression using Eq. (\ref{mmatrixgeneral}), together with the
property $\det(\mathcal{M}) = 1$. We get
\begin{equation}
\Theta (\omega) = 
\frac{|\frac{V_{R}^{-}}{V_{L}^{+}}\mathcal{M}_{12}(\omega) + 1|^{2} + |\frac{V_{R}^{-}}{V_{L}^{+}} - \mathcal{M}_{21}(\omega)|^{2}}
{(1 + \frac{|V_{R}^{-}|^{2}}{|V_{L}^{+}|^{2}})|\mathcal{M}_{22}(\omega)|^{2}}
\end{equation} 
At the singularity frequency point $\omega = \omega_{J}$ and for a generic ratio $V^{R}_{b}/V^{L}_{f}$, the $\Theta (\omega)$-function 
diverges as $\omega \rightarrow \omega_{J}$ and the circuit acts as an amplifier/laser. If on the other hand, we assume that
$V_{R}^{-} = \mathcal{M}_{21}(\omega)V_{L}^{+}$ (perfect adsorbtion condition), we get
\begin{eqnarray} 
\Theta (\omega_{J}) &=
\frac{|\mathcal{M}_{21}(\omega_{J}) \mathcal{M}_{12}(\omega_{J}) + 1|^{2} }
{(1 + |\mathcal{M}_{21} (\omega_{J})|^{2})|\mathcal{M}_{22}(\omega_{J})|^{2}} \nonumber\\
&= \frac{|\mathcal{M}_{22}(\omega_{J}) \mathcal{M}_{11}(\omega_{J})|^{2} }
{(1 + |\mathcal{M}_{21} (\omega_{J})|^{2})|\mathcal{M}_{22}(\omega_{J})|^{2}}=0 
\end{eqnarray} 
In the context of electronics, the two port simultaneous laser/absorber properties are manifest as a delicate balance of the 
driven, marginally stable circuit. The singular behavior of the theoretical $\Theta$ in Fig.~\ref{fig:lasabs}(a), solid curves, 
illustrate that at the Janus frequency $\omega_J$ the injected signals can result in either amplification or complete attenuation, 
depending on the relative amplitude and phase of the injected signals. The perfect absorption condition is particularly sensitive 
to the injection parameters: the deviation of the experimental data, Fig.~\ref{fig:lasabs}(a) dots, is characteristic of component 
imbalance of less than $1\%$. In fact, the minimally absorbing experimental points near the dip in the attenuation curve of 
Fig.~\ref{fig:lasabs}(a) can only be obtained by an independent determination of the minimal reflectance condition at each 
frequency. Fig.~\ref{fig:lasabs}(b) shows this extreme sensitivity to the phase of the right input signal near $\omega_J$ and 
illustrates our current experimental limits to the observation of the Janus condition.


\section{Practical considerations}\label{sec:expcon}

Although the fundamental theoretical aspects of ${\cal PT}$ electronic circuits is straightforward, it is important to realize 
that all physical electronic elements deviate from their ideal intended function in two distinctly different ways. First, all 
have unintentional or stray impedances -- resistive and reactive components -- that can become significant as frequency changes. 
Second, all components, particularly amplifiers, are subject to linearity limits.

The experimental dimer, equivalent to that shown in Fig.~\ref{fig:basic_dimer} with either the inductive or capacitive coupling, consists of a 
pair of coupled $LC$ circuits, one with amplification in the form of the negative resistance, and the other with 
equivalent attenuation. The circuit was shown in Ref. \cite{SLZEK11} to be a simple realization of the ${\cal PT}$-symmetric dimer. 
Each inductor is wound with $75$ turns of \#28 copper wire on $15 cm$ diameter PVC forms in a $6\times6 mm$ loose bundle for an 
inductance of $L = 2.32~mH$. The coils are mounted coaxially with a bundle separation adjusted for the desired mutual inductance $M$. 
The isolated natural frequency of each coil is $\omega_0=1/\sqrt{LC}=2\times10^5s^{-1}$.

The actual experimental circuit includes several additions to that of Fig.~\ref{fig:basic_dimer} acknowledging the physical 
realities mentioned above. First, a resistive component associated with coil wire dissipation is nulled by an equivalent Ohmic 
gain component applied in parallel to each coil. A discussed in section~\ref{sec:elec}, it is not possible to directly apply 
a series Ohmic gain for this compensation. This is our dominant deviation from ideal behavior, however, we have determined 
from simulation that compensating for this series loss by a parallel gain has negligible impact on the ideal ${\cal PT}$ behavior.

Second, additional LF356 op-amps, also used for the negative resistance converter of Fig.~\ref{fig:nics}(a), are used for voltage followers to buffer the voltages $V_{1}$ and $V_{2}$ of Fig.~\ref{fig:basic_dimer}, allowing for a less intrusive capture with the Tektronix DPO2014 oscilloscope used for signal acquisition. 

Finally, small capacitance and gain trims are included to aid in circuit balancing.

Our linearity is constrained by the LF356 op-amps in the negative impedance converters of Fig.~\ref{fig:nics}a. At the $f\sim 30 kHz$ operating frequency, the limits were consistent with the $\pm 12 V$ supply voltage used for the circuit. In fact, the op-amp linearity limited the overall operation frequency of the dimer: higher frequency op-amps are available, but their linearity and input impedance suffer.

The linear nature of our system allows an exact balance of the ${\cal PT}$ symmetry only to the extent that component drift 
over a time scale necessary to perform a measurement is negligible. In the exact phase, real system modes are not perfectly 
achievable: in time, any physical linear system ultimately either shrinks to zero or exponentially grows to the physical 
linearity limit. In the case of our dimer, component precision and drift dictate the accuracy of the ${\cal PT}$ balance to 
approximately $0.1 \%$, and all data was obtained respecting the linearity limits and associated transient time scales.

Experimental practice allows for only a marginal determination gain/loss balance. The chosen  gain/loss parameter $\gamma = 
R^{-1} \sqrt{L/C}$ is set by the loss-side resistance $R$ of Fig.~\ref{fig:basic_dimer}, typically in the range  $1-10 k \Omega$ 
for this work. The gain-side $R$ and capacitance balance are set with the help of the gain and capacitance trims. In the exact 
phase, not too close to $\gamma_{PT}$, the system is trimmed for simultaneous marginal oscillation of both modes with growth or 
decay times greater than $\sim 1 s$, where data is then obtained. The imaginary frequency component is then zero to within $\sim 1 s^{-1}$. 

Very close to the critical point, $\gamma \sim \gamma_{PT}$ attempts to trim the
dimer to the marginal configuration result in either $V=0$ (the gain too small),
or a chaotic interplay of the two modes with the op-amp nonlinearity if the gain is
larger. This behavior serves as one indication that the critical point has been
exceeded. In the vicinity of $\gamma_{PT}$ and beyond, the capacitance trim is kept 
fixed at its asymptotic value, and the gain trim is numerically set to compensate 
for any measured deviation of the gain side $R$ chosen from the desired value. The 
exponential growth or decay rate of transient data obtained then directly gives us 
the imaginary component. Beyond the ${\cal PT}$ point, the exponentially growing mode 
always dominates.  

At this point, these experimental techniques ultimately impact the limits to which the theory is applicable, particularly in the vicinity of the symmetry breaking point where small imbalances can drastically impact the dynamics. We anticipate that, due to the stabilizing nature of resistive loads in the form of transmission lines, the ${\cal PT}$ dimer will provide many opportunities for incorporation into scattering configurations.


\section{Conclusions}\label{sec:concl}
The ${\cal PT}$-symmetric dimer opens a new direction towards investigating novel phenomena and functionalities of ${\cal PT}$-symmetric systems in the {\it spatio-temporal} domain via electronic circuits. This minimal example, which is experimentally simple and mathematically transparent, displays all the universal phenomena encountered in systems with generalized ${\cal PT}$-symmetries. The direct accessibility to all the dynamical variables of the system enables insight and a more thorough understanding of generic ${\cal PT}$-symmetric behavior. In addition, we envision new opportunities for inclusion of ${\cal PT}$ electronics into structures including (nano)-antenna configurations, metamaterials, or microresonator arrays with electronic control over directional signal transmission capabilities and real-time manipulation in the spatio-temporal domain. 

{\it Acknowledgments--}
We acknowledge support by an AFOSR No. FA 9550-10-1-0433 grant, by an NSF ECCS-1128571 grant, and by an Wesleyan Project grant.



\begin{thebibliography}{99}
\bibitem{note1} We use the term ``phase transition" not in the standard thermodynamic sense but rather in the frame of ${\cal PT}$-literature (see for example \cite{BB98}).
\end{thebibliography}
\end{document}